\documentclass[aps,epsfig,preprint,showpacs]{revtex4-1}
\usepackage{amssymb}
\usepackage{graphics}
\usepackage{graphicx}
\usepackage{epsfig}
\usepackage{subcaption}
\usepackage{textcomp}
\usepackage{color}
\usepackage{float}
\begin{document}
\title{Structural Analysis of DNA molecule in a confined shell}
\author{Arghya Maity}
\affiliation{Physics Division, QIC Research Group, Harish-Chandra Research Institute, Prayagraj- 211019, India}
\altaffiliation[Also at: ]{Department of Physics, Birla Institute of Technology \& Science, Pilani - 333 031, Rajasthan, India}
\email{arghyamaityphysics@gmail.com}
\author{Neha Mathur}
\affiliation{Department of Physics, Birla Institute of Technology \& Science, Pilani - 333 031, Rajasthan, India}
\email{p20180045@pilani.bits-pilani.ac.in}
\author{Petra Imhof}
\affiliation{Computer Chemistry Center, Department of Chemistry and Pharmacy, Friedrich-Alexander University (FAU) Erlangen-N\"{u}rnberg, 91052 Erlangen, Germany}
\email{petra.imhof@fau.de}
\author{Navin Singh}
\email{navin@pilani.bits-pilani.ac.in}
\affiliation{Department of Physics, BITS Pilani, Pilani campus-333031, India}


\begin{abstract}
Recent advances in operating and manipulating DNA have provided unique experimental possibilities in many fields of DNA research, especially in gene therapy. Researchers have deployed many techniques, experimental and theoretical, to study the DNA structure changes due to external perturbation. It is crucial to understand the structural and dynamical changes in the DNA molecules in a confined state to understand and control the self-assembly of DNA confined in a chamber or nano-channel for various applications. In the current manuscript, we extend the work  study the effect of confinement on the thermal stability and the structural properties of duplex DNA. The present work is an extension of our previous research works \cite{Maity_EPL_2019, Maity_EBPJ_2020}. For our study we have considered a 1 BNA chain that is confined in a cylindrical geometry. How the geometry of the confinement affects the opening and other structural parameters of DNA molecule is the objective of this manuscript. We have used a statistical model(PBD model) and Molecular dynamics simulations for our purpose. 
\end{abstract}
\maketitle

\section{Introduction}
\label{intro}
Deoxyribonucleic acid(DNA) is an important biomolecule. It has wide applications in the biomedical and biotechnological world. Molecular motor, DNA computer, DNA chip, origami formation, biomedicine, gene therapy are some of the applications which show the importance of this molecule. Degradation of a DNA is major challenge in gene therapy and nanorobotics. The cause of degradation may be a chemical breakdown or sometimes the mechanical forces. In gene therapy research, the performance of the molecule depends on encapsulation as well as effective release of this molecule \cite{Dimitrov_SM_2011}. Many different techniques like complexation with polycations \cite{Putam_Nat_2006}, charged copolymers of different architecture, cationic lipid or liposomes are being used by various research groups. In another technique, researchers confine the DNA molecule within gel \cite{Anne_JCR_2009} or in polymeric nanocapsules(micelles) \cite{Haladjova_SM_2012, Mirkin_AM_2012, Mirkin_Nat_1996}. In general, the short DNA is encapsulated in a spherical inorganic nanoshell of different thicknesses. The nanoshell's thickness and DNA integrity at a higher temperature are crucial parameters in handling the encapsulation process. Due to the restricted behaviour of DNA molecule during encapsulation, it is of deep interest to investigate the tube diameter's effect on the DNA molecule's stability. The confinement restricts the conformation and movement of DNA molecules in the cell due to which there is substantial changes in the thermodynamic properties of DNA molecules \cite{Sanjay_EPL_2017, Turner_PRL_2002, Maity_EBPJ_2017, Akabayov_NatC_2013}. The DNA packing in eukaryotic chromosomes, viral capsids in the confined space are some examples that show the importance of the studies related to DNA in confined geometry. How the confinement affects the structural parameters of DNA is an interesting sbuject. Motivated by all these different kinds of experimental and simulation based studies in different geometries, we investigate the effect of confinement on the stability and structural changes of DNA in the current manuscript. We present the work in the following sections. First we discuss the results on melting of DNA using the PBD model. We then discuss the results obtained using Gromacs package for the confined DNA. The paper conclude with a discussion on the obtained results.

\section{Methods}
\label{methods}

To investigate the thermodynamic properties of DNA in a confined shell, we choose the Peyrard-Bishop-Dauxois (PBD) model. The model is quasi-one-dimensional and describes the motion of the molecules through the stretching of the hydrogen bonds between the bases in a pair \cite{Dauxois_PRE_1993}. The model underestimates the entropy associated with the different conformations of the molecule \cite{Frank_PLR_2014}. Since the base pairs in the linear form of PBD model are allowed to move along $y$-axis only, the entropic contributions from the overall motion of bases are ignored. The bases in open state are having more degree of freedom than predicted by this model. In the linear form of the model effect of the helicoidal nature of the molecule is also ignored. However, there have been several attempts to extend the model to study the circular and twist forms of DNA in thermal as well as force ensembles \cite{Cocco_PRL_1999,Zoli_JCP_2021}. Several researchers have extended the model to three-dimensions \cite{Zoli_PCCP_2019, Weber_CPL_2020}. The PBD model has been extended to study the effect of molecular crowders as well as the effect of confinement on the thermal stability of DNA molecule \cite{Zoli_PCCP_2019, Maity_EPL_2019, Maity_EBPJ_2020}. The mathematical form of the model that contains $N$ base pairs, is as follows:
\begin{equation}
\label{eqn1}
H = \sum_{i=1}^N\left[\frac{p_i^2}{2m}+ V_m(y_i) \right] + \sum_{i=1}^{N-1}\left[V_s(y_i,y_{i+1})\right],
\end{equation}
here $y_i$ represents the stretching from the equilibrium position of the hydrogen bonds. The momentum term of the Hamiltonian is $p_i = m\dot{y}_i$. The reduced mass, $m$, is considered the same for both $AT$ and $GC$ base pairs. The interaction between the nearest base pairs along the chain, the stacking interaction, is 
$$ 
V_s(y_i,y_{i+1}) = \frac{k}{2}(y_i - y_{i+1})^2[1 + \rho e^{-b(y_i + y_{i+1})}].
$$
\begin{figure}
\begin{center}
\includegraphics[height=1.0in,width=2.3in]{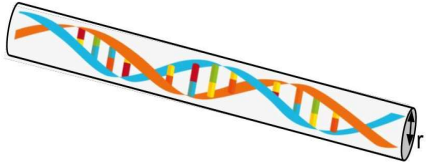}
\caption{\label{fig01} \small The schematic representation of the DNA molecule in a confined cell of cylindrical geometry. The $r$ is the distance of the confined wall from the DNA strand. So the diameter of the cylinder becomes $r$+DNA diameter.}
\end{center}
\end{figure}
The sequence heterogeneity affects the stacking interaction along the strand. This can be taken care of through the single strand elasticity parameter $k$. One can take the variable $k$ according to the distribution of bases along the strand. However, for the current investigation we have taken a constant value of $k$. The anharmonicity in the strand elasticity is represented by $\rho$. The parameter, $b$, describes the range of anharmonicity. For our studies, we choose model parameters $k$ = 0.015 eV\AA$^{-2}${}, $\rho = 5.0$, and $b = 0.35$  \AA$^{-1}${} \cite{Maity_EBPJ_2020}. The value of the parameter $\rho$ defines the sharpness in the transition from double-strand to single-strand \cite{Navin_epje_2005}. The Morse potential represents the hydrogen bonding between the two bases in the $i^{\rm th}$ pair is: 
$$
V_m (y_i) = D_i(e^{-a_iy_i} - 1)^2.
$$
Here $D_i$ represents the potential depth, and $a_i$ represents the inverse of the width of the potential well. The dissociation energy, $D_i$, is a representation of the hydrogen bond energy that binds the $AT$ and $GC$ pairs while $a_i$ represents bond stiffness. From previous results, we know that the bond strengths of these two pairs are in an approximate ratio of $1.25$ - $1.5$ as the GC pairs have three while AT pairs have two hydrogen bonds \cite{Weber_CPL_2020, Zoli_PCCP_2019}. The potential parameters are taken as $a_{AT} = 4.2 \; {\rm \AA^{-1}},a_{GC} = 1.5\times a_{AT}$ and $D_{AT}$ = 0.055 eV while, $D_{GC} = 1.5\times D_{AT}$. We can study the thermodynamics of the transition by evaluating the partition function. For a sequence of $N$ base pairs, the canonical partition function can be written as:
\begin{equation}
\label{eqn3}
Z = \frac{1}{h}\int \prod_{i=1}^{N}\left\{dy_idp_i\exp(-\beta H)\right\} = Z_pZ_c,
\end{equation}
where $Z_p$ corresponds to the momentum part of the partition function while the $Z_c$ contributes as the configurational part of the partition function. The factor $1/h$ is included to take care of dimensionallity of the partition function. The momentum part is $(2\pi mk_BT)^{N/2}$. The configurational part of the partition function, $Z_c$, is defined as \cite{Peyrard_PRL_1989}, 
\begin{equation}
\label{eqn4}
Z_c = \int \left[\prod_{i=1}^{N-1} dy_i  K(y_i,y_{i+1})\right]dy_NK(y_N)
\end{equation}
where 
$ K(y_i,y_{i+1}) = \exp\left[-\beta H_c(y_i,y_{i+1})\right]$ and $K(y_N) = \exp\left[-\beta V(y_N)\right]$. For the homogeneous chain, one can evaluate the partition function using the transfer integral (TI) method by applying the periodic boundary condition. For a heterogeneous sequence, the calculation of the partition function is not straightforward. In a heterogeneous sequence, the $i^{th}$ site may not be of the same nature as its $i-1$ and $i+1$ sites. Researchers have used methods like quench disorder, or extended transfer matrix (ETMA) to study the melting of a heterogeneous DNA molecule \cite{Cule_PRL_1997,Zhang_PRE_1997}. With open boundary condition, the matrix multiplication method can address the calculation of partition function  \cite{Campa_PRE_1998}. To avoid the divergence of the partition function, there is a need of proper cut-off for the integral appearing in Eq.(\ref{eqn4}). van Erp et al. have shown that the upper cut-off as $\approx 144 \; {\rm \AA}$ \cite{Erp_epje_2006} is sufficient to avoid the divergence. Dauxois and Peyrard have shown that $T_m$ converges rapidly with the upper limit of integration \cite{Dauxois_PRE_1995}. We calculate $T_m$ for different values of upper cut-off and found that the choice of 200 ${\rm \AA}$ is sufficient to avoid the divergence of the partition function. The configurational space for our calculations extends from -5 ${\rm \AA}$ to 200 ${\rm \AA}$. To calculate the partition function, we generate matrices using Eq.(\ref{eqn4}) and multiply the obtained matrices one by one. After the proper cut-offs, the next task is to discretize the integral. To get a precise value of melting temperature ($T_m$) we have observed that Gaussian quadrature is the most effective quadrature. We have found that discretization of the space with 900 points is sufficient to get an accurate value of $T_m$ \cite{Amar_PRE_2015}. Once we evaluate the partition function, we can determine the thermodynamic quantities of interest by evaluating the Helmholtz free energy of the system. The Helmholtz free energy ($f$) and specific heat ($C_v$) per base pair are: 
$$
f(T) = - \frac{k_B T}{N}\ln Z, \qquad\qquad C_v = -T(\partial^2 f/\partial T^2)
$$
The melting temperature ($T_m$) of the chain is evaluated from the peak in the specific heat. 

\section{Melting of 1 BNA}
\label{melting}

To study the thermodynamics of DNA in the cylindrical confinement we adopt the scheme proposed in our earlier work \cite{Maity_EBPJ_2020}. Due to the surrounding cellular environment, the movement of base pairs will be restricted, which in turn affects the overall movement of the molecule. We restrict the configuration space of the system, as shown in Fig. \ref{fig01}. The lower limit of integration is -5 ${\rm \AA}$ while the upper limit of integration for each base pair is $r$. The $r$ is the distance of the confined wall from the DNA strand. Using the modified scheme we calculate the partition function and hence evaluate all the thermodynamical properties of DNA confined in a cylindrical shell.
\begin{figure*}[hbt]
\begin{center}
\includegraphics[height=2.0in,width=4.50in]{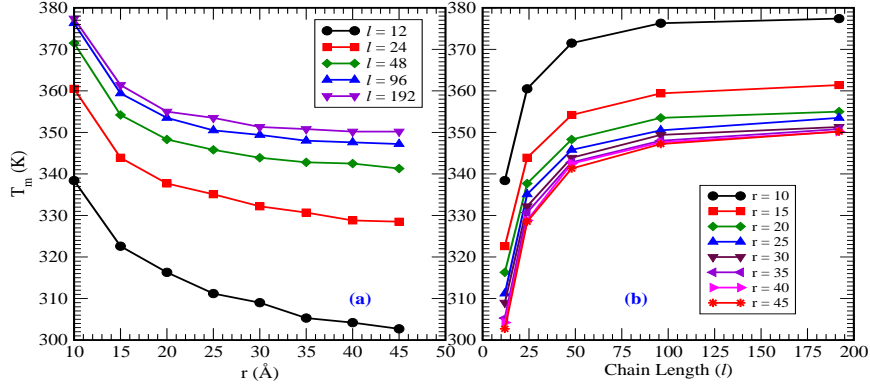}
\caption{\label{fig02}\small The plot showing the change in the melting temperature of the DNA chain of different lengths. (a) The variation in the melting temperature with the increasing cylinder radius for different chains. (b) The change in the melting temperature with the increasing chain length.}
\end{center}
\end{figure*}
\begin{figure*}[hbt]
\begin{center}
\includegraphics[height=2.in,width=4.5in]{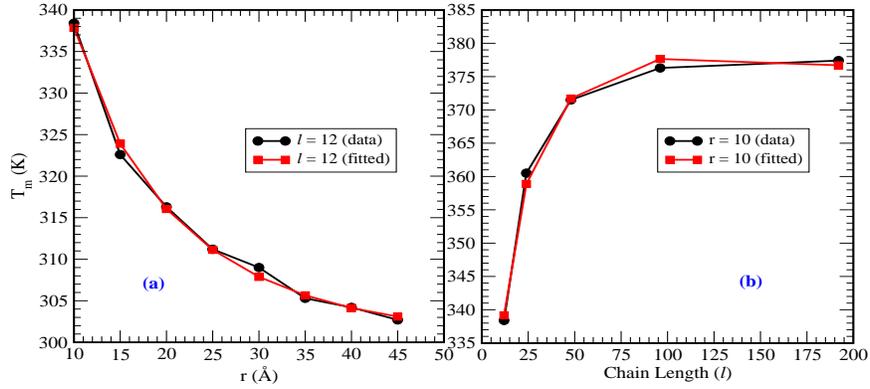}
\caption{\label{fig03}\small (a) The variation in melting temperature with the radius of cylindrical shell for 12-bp chain. The best fit parameters are: $T_m^0 = 481.43$ K, $\lambda_1 = -86.08,\; \& \; \lambda_2 = 10.31$. (b) The variation in the melting temperature with the chain length for the DNA confined in a cylinder shell of radius 10 \AA{}. The best fit parameters are: $T_m^0 = 211.93$ K, $\lambda_1 = 69.01, \; \& \; \lambda_2 = -7.17$.}
\end{center}
\end{figure*}
In the first part of the study, we consider a DNA molecule of 12 base pairs confined in a cylinder. Using the PBD model, we calculate the melting temperature of the chain. Our interest is to find out a correlation between the chain size and the radius of the cylinder with the melting temperature of the chain. We first calculate the melting temperature of a chain of 12 base pairs. The melting temperature ($T_m$) of the chain that is confined in the cylinder of radius 10 \AA{} is 338.4 K. We increase the radius of the cylinder up to 45 \AA{} and find the melting temperature of the molecule. The $T_m$ for 12 bp chain in a cylinder of 45 \AA{} is 302.7 K. To see the effect of chain length on the melting temperature, we double the chain length and increases it up to 192 base pairs. For all the chains, one typical pattern we observe. There is a decrease in the melting temperature, which saturates to a particular value. The interesting part of the outcome is the pattern of the decay in the $T_m$ with increasing size of cylinder (in terms of radius). To indentify the pattern we fit the obtained results with a non-linear cuver fitting program. The best fit equation is:
\begin{equation}
\label{eqn8}
T_m = T_m^0 + \lambda_1\ln(x) + \lambda_2\ln^2(x)
\end{equation}
where $x$ is either the number of base pairs ($l$) in a given chain (for fixed cylinder width) or the radius ($r$) of the cylindrical shell (of fixed DNA chain). The parameters, $T_m^0, \; \lambda_1, \; \& \; \lambda_2$ are fitting parameters. 

We find that the decay in the $T_m$ with radius of the cylinder is logarithmic in nature. For the chain of 24 base pairs, it saturates to 328.8 K, for 48 base pairs 341.3 K, while for 192 base pairs chain it saturates 350.2 K. To understand more about the stability of the molecule that is confined in the cylinder fix the radius of the cylinder and vary the length of the DNA molecule. With the increase in the chain length, we observe that the melting temperature of the chain increases with the increase in the chain length and saturates to a particular value. For the DNA molecule of 12 base pairs, that is confined in a cylindrical shell of radii 10 \AA{} to 45 \AA{}, we find the best fit parameters as $T_m^0 = 481.43$ K, $\lambda_1 = -86.08,\; \& \; \lambda_2 = 10.31$. Please note that $\lambda_1$ is negative here. When we confine DNA molecule of different lengths in a cylinder of radius 10 \AA{}, we find the best fit values as $T_m^0 = 211.93$ K, $\lambda_1 = 69.01, \; \& \; \lambda_2 = -7.17$. Here, the value of $\lambda_1$ is positive. The third term in the Eq. (\ref{eqn8}) is the correction term. We also find the best fit values for other confined radii, however, to avoid overcrowding of plots we are not showing the results here. We show the best fit plots for both the cases through the fig. \ref{fig03}.

\section{The RMSD value of 1 BNA}

\begin{figure}
\begin{center}
\includegraphics[height=2.25in,width=3.5in]{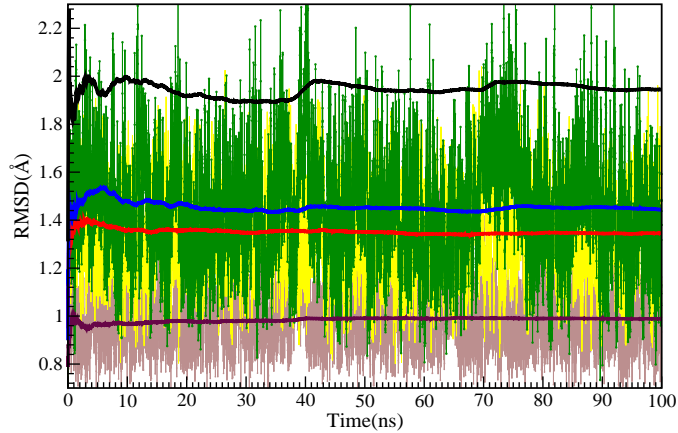}
\caption{\label{fig04}\small RMSD calculation results are plotted for three cylindrical radii ($r$). The running average is shown for each RMSD fluctuations and for not confined DNA system also. The brown line is for $r$ = 11.5 \AA{}, the red line is $r$ = 12.5 \AA{}, the blue line is for $r$ = 15.0 \AA{}, and the black line is for unconfined DNA.} 
\end{center}
\end{figure}

The next part of the study focuses on the changes in the average dimension of DNA molecule that is confined in a cylinder. We also investigate the effect of confinement on the structural parameters of the DNA molecule. We choose 1 BNA (PDB code) confined in a cylindrical shell. Using {\it Gromacs 4.0.4} package and choosing AMBER14sb$\_$parmbsc1 as a force field we study the DNA in confined geometry. The 1 BNA is placed in a cubical box such that it is 1.0 nm apart from edge of the box. The dimension of the box is (6.44747; 6.44747; 6.44747). We use spc216 model, which is a generic equilibrated 3-point solvent model and add $0.1M$ of $NaCl$ in the solution. In the simulation we restrict every base pair according to the cylindrical geometry. For the geometry, the radius of the cylinder ($r$) is the fundamental parameter (see Fig.\ref{fig01}). Before taking any measurement, we have to ensure that the solvent and ions around the DNA are well equilibrated. The minimization ensures that the system is at equilibrium, in terms of geometry and solvent orientation. Sometimes the solvent is optimized within itself and does not optimize with the solute. The solvent needs to be brought to the temperature we wish to simulate and establish the proper orientation about the solute (the DNA). Once we arrive at the equilibrium temperature (based on kinetic energies), we apply pressure on the system until it reaches the proper density. We change the reference temperature so that it takes care of the system's relaxation time and equilibrate the system as per the standard protocol, first in the NVT ensemble then in the NPT ensemble. Once we achieve a proper equilibration, we are ready to understand the evolution of the system with time. For data collection, we release the position restraints and run the simulation. The cutoffs for short-range electrostatic interactions and short-range van der Waals are chosen as 1.4 nm. Here we use a modified Berendsen thermostat for our simulation. We run the MD simulation for 100-ns and use \textit{trjconv} as a tool to strip out coordinates, the periodicity, time units, and frame frequency. By varying the radius of the cylinder ($r$) from $11.5 - 15.0$ \AA{} we calculate the RMSD value for each radius. As can be seen in Fig.\ref{fig04}, the flexibility of the DNA, as manifested by the running average of its  RMSD, is impacted by the confinement. The RMSD values are smallest for the most confined situation, i.e. cylinder radius of 11.5 \AA{}, the RMSD is  $\sim 0.99 \pm 0.054$ \AA{} and larger for a confinement radius of 15.5 \AA, the RMSD value is $1.45 \pm 0.171$ \AA{} which are significantly below the RMSD values computed for unconfined DNA ($\sim 1.95 \pm 0.052$ \AA{}). 

\section{Structural parameters of a confined DNA}
\label{simulate}

To analyze different structural features of DNA, we take the help of \textit{curves+}. The \textit{curves+} is the revised version of the \textit{curves} approach \cite{Lavery_NAR_2009} which is used to analyze the molecular dynamics trajectories and generates time series. It follows the international conventions for nucleic acid analysis. The \textit{curves+} provides the standard geometrical figure of DNA bases in a data file by analyzing molecular dynamics trajectories and provides all helical and backbone parameters \cite{Wilma_JMB_2001, Bansal_CAB_1995}. To describe the system's parameters, we can either evaluate about the local helix axes between two base pairs or an overall helix axis of the molecule. The parameters obtained in these two cases, however, differ significantly. The \textit{curves+} avoids the issues related to the local and global helical parameters. The \textit{curves+} set its reference system using chosen base atoms. There are \textit{C1'}, \textit{N1(Y)/N9(R)}, and \textit{C2(Y)/C4(R)} in standard bases where $Y$ represents a pyrimidine base, and $R$ represents a purine base. To understand the underlying mathematics, please see the references \cite{Lavery_NAR_2009}. With this tool, we can produce a complete set of helical parameters. We study the complete set of helical parameters in the current work, including translational and rotational parameters. 

We simulate 100 ns and analyze the changes in the geometrical structure of DNA molecule for two different radii 11.5 \AA {} and 15.0 \AA{} of the cylinder through curves+. The intra-base pairs parameters hold three translational parameters - \textit{shear}, \textit{stretch}, \textit{stagger}, and three rotational parameters - \textit{buckle}, \textit{propeller}, \textit{opening}. The zero values of these parameters represent canonical Watson-Crick base pairs, and the non-zero values show the distortion of the short and long axis of the base pairs. We calculate the parameters using the rigid-body transformation. Here we are considering a short DNA molecule; the end effects may change the average value of RMSD. We execute the MD simulation for 100 ns to ensure equilibration, which puts more flexibility on the terminal base pairs. In the past, researchers have deployed several techniques to reduce the end effect in simulations \cite{Bevan_BJ_2000, Dhananjay_JCP_2009, Cheng_NAR_2006, Dai_PRL_2008, Long_JPC_2006}. One of the simple ways among them is to focus on the structural changes of base pairs in the centre of the chain, assuming that they are unaffected by the end's fluctuations. The \textit{curves+} provides the average value of base-pair parameters (intra, inter, and BP-axis parameters). We consider a chain of 12 bps, and to avoid the end effect distortion, we prefer to have an average value of 4 bp to 9 bp for \textit{intra}-base pair and \textit{base pair-axis} parameters. The \textit{inter-base pair} parameters are pair junction parameters; hence the average values neglect the first two and last two base pair effects.

\subsection{Effect on intra-base pair parameters}

Let us discuss the effect of confinement on the rotational as well as translational intra-base parameters. After 100 ns of simulation, we find the stagger value of free DNA (unconfined) as $13.82$ \AA{}. The large value shows that at 340 K, the hydrogen bonds of DNA are broken. When the DNA molecule is confined in a cylinder of radius $r = 15.0 \; {\rm \AA}$, the stagger value decreases to -0.02 \AA{}. The suppression in the translational parameters is due to the restricted motion of the base pairs. Similarly, there is a substantial decrease in the values of rotational parameters: {\it propeller} and {\it opening}. It is known that the Watson-Crick base pairs hold a negative value of {\it propeller}. The value of {\it propeller} of free DNA is found to be 24.3 after a run of 100 ns simulation.  When the DNA is confined in a cylinder of radius $r = 15.0 \; {\rm \AA}$, the value of {\it propeller} decreases to -5.80, which confirms that now DNA is in the double-stranded state. The confinement does not allow the chain to denature, which is reflected through the negative propeller angle. The value of {\it opening} of free DNA is 11.91, it decreases to 1.4 for $r = 15.0 \; {\rm \AA}$ which further decreases to 1.08 for $r = 11.5 \; {\rm \AA}$. The shorter the radius, the larger the restriction on the motion of the DNA molecule. The suppression in these parameters attributes to the stability of the structure of the molecule.
\begin{table*}
\caption{\small The average value of intra-base pair parameters of the unconfined and confined DNA. In the last row, we show the values obtained by Lavery et al \cite{Lavery_NAR_2009}. Their values are at 300 K while our values are at 340 K.}
\label{tab1}
\begin{center}
\scalebox{0.75}{
\begin{tabular}{|c|c| c | c | c | c | c|c| }\hline
\multicolumn{7}{|c|}{Intra-BP Parameters}\\ \hline
Parameters & Shear(\AA) & Stretch(\AA) & Stagger(\AA) & Buckle($^0$) & Propeller($^0$) & Opening($^0$)\\ \hline
Unconfined & 0.01 & -14.46 & 13.82 & -7.3 & 24.3 & 11.91  \\ \hline
$r=15.0$ \AA & 0.02 & 0.02 & -0.02 & 2.05 & -5.80 & 1.40 \\ \hline
$r=11.5$ \AA & -0.07 & 0.02 & -0.10 & -2.35 & -9.26 & 1.08 \\ \hline
From refs. \cite{Lavery_NAR_2009} & -0.04 & -0.17 & 0.21 & 0.3 & -13.7 & 1.0  \\ \hline
\end{tabular}
}
\end{center}
\end{table*}

\begin{figure}
\begin{center}
\includegraphics[height=2.25in,width=1.5in]{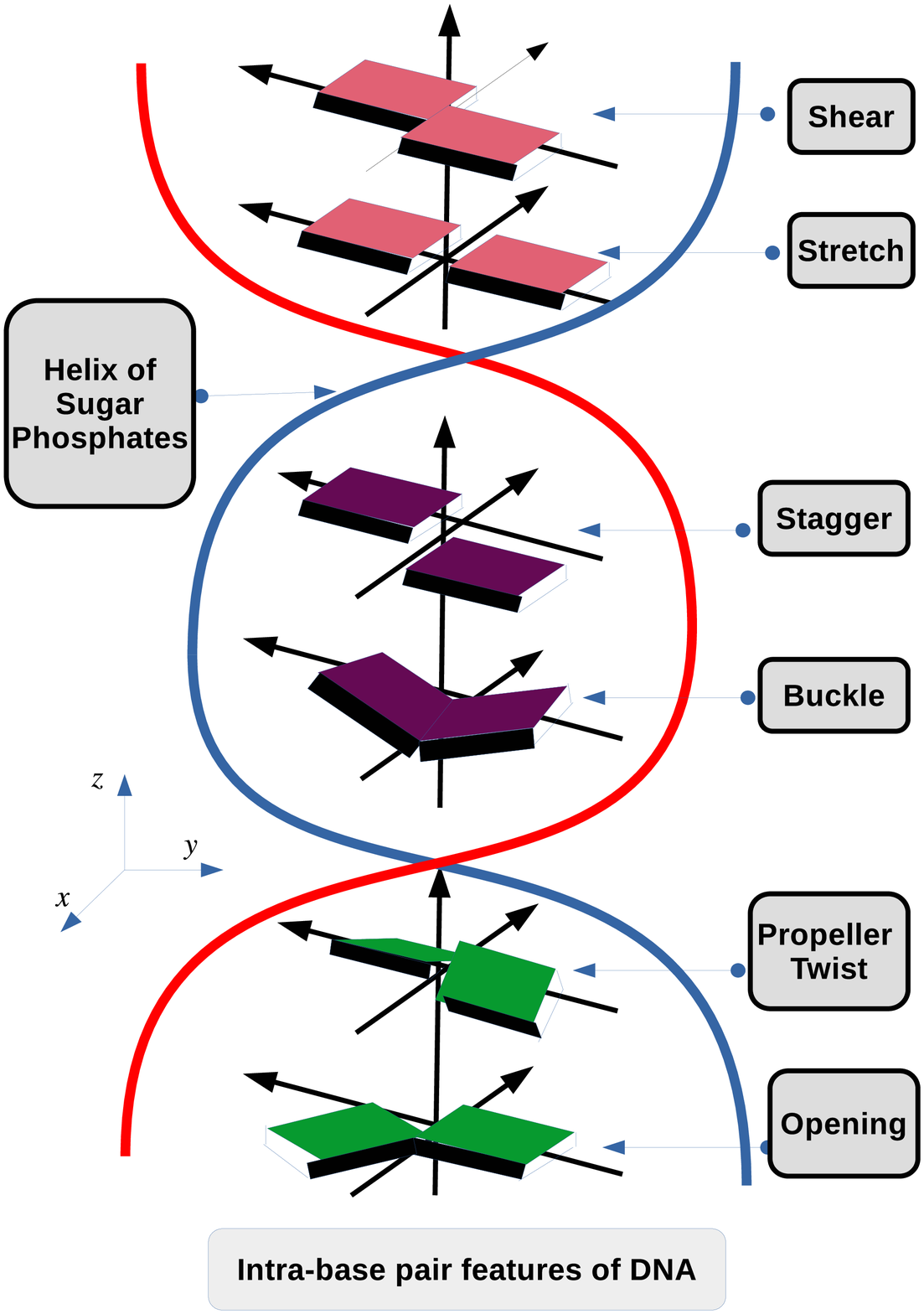}
\includegraphics[height=2.25in,width=1.5in]{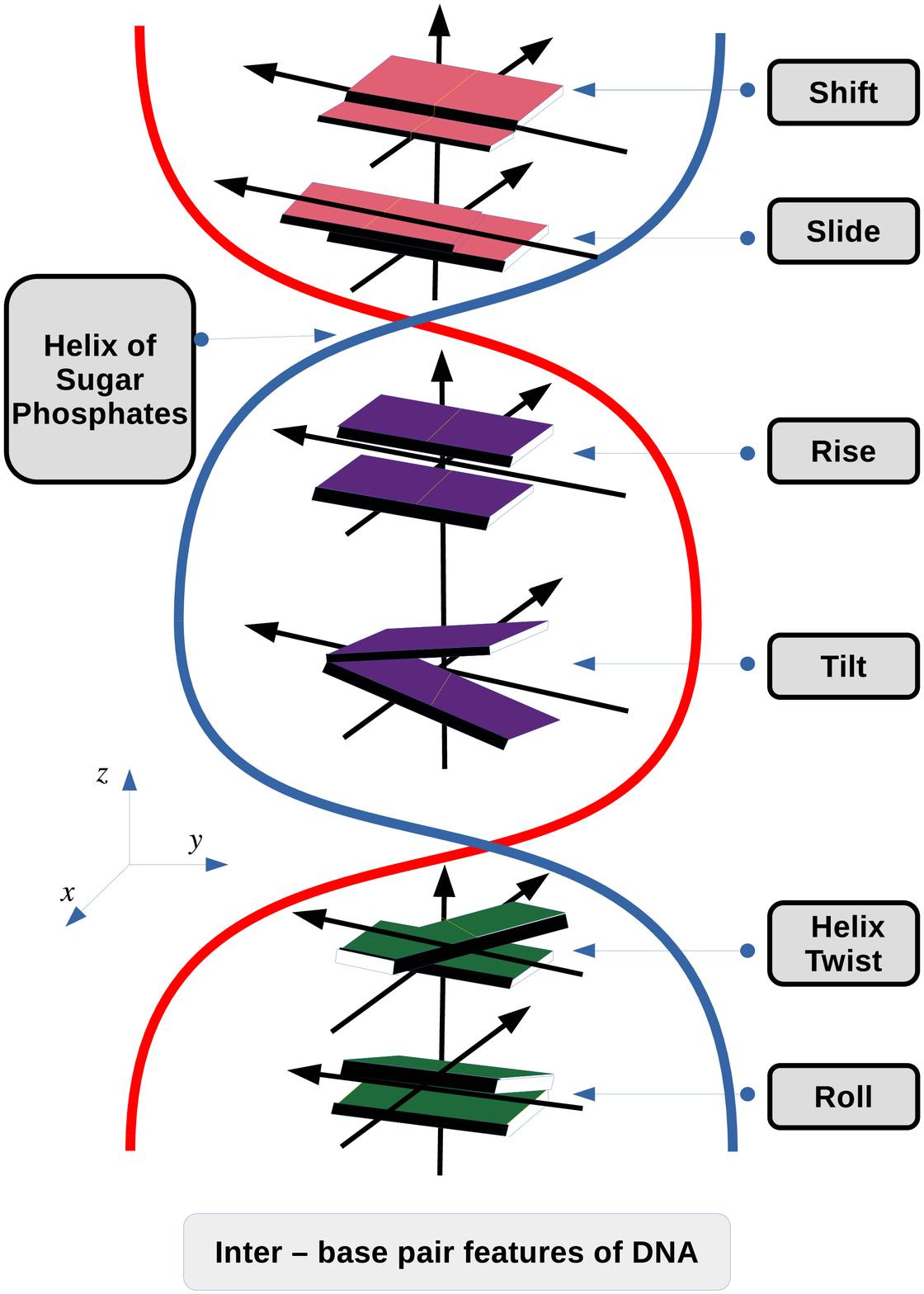}
\end{center}
\caption{\label{fig56}\small The schematic of intra and inter base pair orientations in DNA molecule.} 
\end{figure}

\subsection{Effect on inter-base pair parameters}

The {\it inter-base pair} parameters also hold three translational parameters- {\it shift, slide, rise}, and three rotational parameters - {\it tilt, roll, twist}. The translational and rotational parameters attribute to two successive base-pairs relative positions to their short axes, long axes, and normals. In literature, these are also called pair junction parameters. Reference frames are defined for the two base pairs, and each base pair frame is referred to as the mid-frame that ion Rodrigues' formula. Please see the refs \cite{Lavery_NAR_2009} and fig.\ref{fig56} for more details. 
\begin{table*}
\caption{\small The average value of inter-base pair parameters of the unconfined and confined DNA. In the last row, we show the values obtained by Lavery et al \cite{Lavery_NAR_2009}. Their values are at 300 K while our values are at 340 K.}
\label{tab2}
\begin{center}
\scalebox{0.9}{
\begin{tabular}{|c|c| c | c | c | c | c| }\hline
\multicolumn{7}{|c|}{Inter-BP Parameters}\\ \hline
Parameters & Shift(\AA) & Slide(\AA) & Rise(\AA) &Tilt($^0$) &Roll($^0$) &Twist($^0$)\\ \hline
Unconfined & 2.24 & 6.77 & 20.28 & -17.20 & 5.91 & -17.00  \\ \hline
$r=15.0$ \AA & 0.14 & -0.17 & 3.24 & 1.45 & 1.27 & 34.50 \\ \hline
$r=11.5$ \AA & 0.10 & 0.06 & 3.30 & 0.34 & 0.11 & 36.21 \\ \hline
From refs. \cite{Lavery_NAR_2009} & -0.02 & 0.14 & 3.36 & -0.2 & -0.3 & 35.8  \\ \hline
\end{tabular}
}
\end{center}
\end{table*}
The changes in the inter-base parameters due to the confinement is shown in Table-\ref{tab2}. In general, the {\it twist angle} in 1 BNA is found to be in between $20^0$ and $40^0$. After a simulation of 100 ns, the twist angle in free DNA is found to deviate from the mentioned range. When we confine the DNA in a cylinder of radius $r = 15 \; {\rm \AA}$, {\it twists} angle is $34.50$, which shows that now the DNA is in the double helix structure. The {\it roll} angle generally lies between $0^0$ and $15^0$ in 1 BNA structures \cite{Olson_PNAS_1998}. The {\it roll} and {\it twist} angles are associated with the changes in the translational parameter slide. 

\subsection{Effect on base-pair axis parameters}

To make the analysis conclusive, we investigate the effect of confinement on the base-pair axis parameters (see table-\ref{tab3}) of the DNA molecule. Again we have two translational parameters- {\it xdisp, ydisp}, and two rotational parameters - {\it inclination, tip}. The {\it xdisp} tells us the movement of bases towards the grooves, while the {\it ydisp} tells us the movement of bases perpendicular to the grooves. The rotational parameters, the {\it inclination} shows the rotation around the short axis of base pairs, while the {\it tip} shows the rotation around the long axis of the base pairs.
\begin{table}
\caption{\small The average value of base-pair axis parameters of unconfined and confined DNA. In the last row, we show the values obtained by Lavery et al \cite{Lavery_NAR_2009}. Their values are at 300 K while our values are at 340 K. }
\label{tab3}
\begin{center}
\scalebox{0.9}{
\begin{tabular}{|c|c| c | c | c |c| c|}\hline
\multicolumn{5}{|c|}{Base pair-axis Parameters}\\ \hline
Parameters & x-disp(\AA) & y-disp(\AA) & inclin($^0$)& tip($^0$)\\ \hline
Unconfined & -5.22 & 3.12 & -16.98 & -69.03   \\ \hline
$r=15.0$ \AA & -0.29 & -0.59 & -0.10 & -0.60 \\ \hline
$r=11.5$ \AA & -0.24 & -0.10 & 0.45 & -1.01 \\ \hline
From refs. \cite{Lavery_NAR_2009} & 0.27 & 0.11 & -0.1 & -1.0   \\ \hline
\end{tabular}
}
\end{center}
\end{table}
The change in the values of all the base-pair axis parameters of free DNA after 100 ns of simulation shows that DNA at 340 K is no more in a helical state. When the molecule is confined in the cylinder, we find that even after a run of 100 ns, the base pair axis parameters lies in the range that confirms the double-stranded conformation of the DNA molecule. All three parameters (intra, inter and base-pair axis) indicate that while the free DNA at 340 K after 100 ns is in the unzipped state, the confined geometry's DNA molecule is still in a zipped state. 

\section{Summary}
\label{summ}

In the present work, we have considered the DNA molecules that is confined in a cylindrical shell. In the first part of the study, we have considered heterogeneous DNA molecules of different lengths (12-192 bps) and have calculated the melting temperature of the system. The main objective of this part of the study is to find out a correlation between the melting temperature ($T_m$) of the system with the size of the cylinder as well as with the chain size. For all the studies, we have considered cylinder of infinte length and varied the radius of the cylinder. We have used a non-linear curve fitting equation and have found that the $T_m$ varies logarithmically with the chain length as well as the radius of the cylinder. The melting temperature, $T_m$, increases with system size while it decreases with increasing cylinder radius.  The question which we were interested to investigate is there any correlation between the decrease in the melting temperature with the decrease in the radius of the cylinder. In the next part of the study, we have used {\it Gromacs}  package, version 4.0.4, to study the change in the average size of the 1 BNA molecule with the changing size (radius) of the cylinder. With the change in the cylinder's size (radius only), we have calculated the RMSD (root mean square deviation) values of the confined DNA at different temperatures. The RMSD calculations showed that the molecule is more stable when it is placed inside a cylinder. The results are trivial and well in agreement with the first part of the current investigation and our earlier work \cite{Maity_EPL_2019, Maity_EBPJ_2020}. The reason for the stability is as follows: the decrease in the cylinder's radius, the confining wall becomes closer, which suppresses the molecule's entropy and hence the RMSD value decreases. Our main objective of the current work was not only to the study the of stability of the molecule in a confined shell but also to study the effect of confinement on the various structural parameters of the DNA molecule. That is the reason for the last part of the investigation. Here, we have calculated all the helical and backbone parameters of 1 BNA molecule confined in a cylindrical shell. We have run the 100 ns simulation, and using the \textit{curve+} tool, we have done a systematic analysis of the change in the intra and inter base-pair parameters of 1 BNA molecules in confined shell. Due to the restricted space there is a substantial change in the various structural parameters of the molecule. We have compared our findings with the findings of Lavery {\it et al} \cite{Lavery_NAR_2009}. Our calculations of inter as well intra base pair parameters of confined DNA at 340 K are close to the unconfined DNA at 300 K found by Lavery {\it et al} \cite{Lavery_NAR_2009}. Although there are certain parameters which are not in good agreement. The reason for mismatch is obvious. The molecule that is under confinement even at 340 K, although is more stable, all the inter as intra base pairs parameters can not be same. 

The current investigation presents a systematic (probably first) analysis of the effect of confinement not only on the thermodynamics of the molecule but also on the changes in microscopic parameters, inter and intra base-pair parameters, of DNA molecules. We all know that the cell is crowded and there is a very finite space available to the biomolecules. From biomedical point of view, we know that DNA is confined in a shell to protect it from any damage. Our studies may enhance our understanding of the molecule that is confined in a shell, the process used in the area of gene therapy. As a future study, we would like to study conical geometry's effect on the structural changes in the DNA molecule.

\section{Acknowledgments}

We acknowledge the financial support from the Department of Science and Technology, New Delhi (EMR/2017/002451). Arghya Maity acknowledges the Free University of Berlin for the scholarship and the lab facility for the research work.  We would like to thank Mr. Sachin Mishra, an undergraduate student at BITS Pilani to help us in the analysis using {\it curve+}.

\bibliography{ms_simul_arxiv} 

%merlin.mbs apsrev4-1.bst 2010-07-25 4.21a (PWD, AO, DPC) hacked
%Control: key (0)
%Control: author (8) initials jnrlst
%Control: editor formatted (1) identically to author
%Control: production of article title (-1) disabled
%Control: page (0) single
%Control: year (1) truncated
%Control: production of eprint (0) enabled
\begin{thebibliography}{35}%
\makeatletter
\providecommand \@ifxundefined [1]{%
 \@ifx{#1\undefined}
}%
\providecommand \@ifnum [1]{%
 \ifnum #1\expandafter \@firstoftwo
 \else \expandafter \@secondoftwo
 \fi
}%
\providecommand \@ifx [1]{%
 \ifx #1\expandafter \@firstoftwo
 \else \expandafter \@secondoftwo
 \fi
}%
\providecommand \natexlab [1]{#1}%
\providecommand \enquote  [1]{``#1''}%
\providecommand \bibnamefont  [1]{#1}%
\providecommand \bibfnamefont [1]{#1}%
\providecommand \citenamefont [1]{#1}%
\providecommand \href@noop [0]{\@secondoftwo}%
\providecommand \href [0]{\begingroup \@sanitize@url \@href}%
\providecommand \@href[1]{\@@startlink{#1}\@@href}%
\providecommand \@@href[1]{\endgroup#1\@@endlink}%
\providecommand \@sanitize@url [0]{\catcode `\\12\catcode `\$12\catcode
  `\&12\catcode `\#12\catcode `\^12\catcode `\_12\catcode `\%12\relax}%
\providecommand \@@startlink[1]{}%
\providecommand \@@endlink[0]{}%
\providecommand \url  [0]{\begingroup\@sanitize@url \@url }%
\providecommand \@url [1]{\endgroup\@href {#1}{\urlprefix }}%
\providecommand \urlprefix  [0]{URL }%
\providecommand \Eprint [0]{\href }%
\providecommand \doibase [0]{http://dx.doi.org/}%
\providecommand \selectlanguage [0]{\@gobble}%
\providecommand \bibinfo  [0]{\@secondoftwo}%
\providecommand \bibfield  [0]{\@secondoftwo}%
\providecommand \translation [1]{[#1]}%
\providecommand \BibitemOpen [0]{}%
\providecommand \bibitemStop [0]{}%
\providecommand \bibitemNoStop [0]{.\EOS\space}%
\providecommand \EOS [0]{\spacefactor3000\relax}%
\providecommand \BibitemShut  [1]{\csname bibitem#1\endcsname}%
\let\auto@bib@innerbib\@empty
%</preamble>
\bibitem [{\citenamefont {Maity}\ \emph {et~al.}(2019)\citenamefont {Maity},
  \citenamefont {Singh},\ and\ \citenamefont {Singh}}]{Maity_EPL_2019}%
  \BibitemOpen
  \bibfield  {author} {\bibinfo {author} {\bibfnamefont {A.}~\bibnamefont
  {Maity}}, \bibinfo {author} {\bibfnamefont {A.}~\bibnamefont {Singh}}, \ and\
  \bibinfo {author} {\bibfnamefont {N.}~\bibnamefont {Singh}},\ }\href
  {\doibase 10.1209/0295-5075/127/28001} {\bibfield  {journal} {\bibinfo
  {journal} {{EPL} (Europhysics Letters)}\ }\textbf {\bibinfo {volume} {127}},\
  \bibinfo {pages} {28001} (\bibinfo {year} {2019})}\BibitemShut {NoStop}%
\bibitem [{\citenamefont {Maity}\ and\ \citenamefont
  {Singh}(2020)}]{Maity_EBPJ_2020}%
  \BibitemOpen
  \bibfield  {author} {\bibinfo {author} {\bibfnamefont {A.}~\bibnamefont
  {Maity}}\ and\ \bibinfo {author} {\bibfnamefont {N.}~\bibnamefont {Singh}},\
  }\href {\doibase 10.1007/s00249-020-01462-9} {\bibfield  {journal} {\bibinfo
  {journal} {European Biophysics Journal}\ }\textbf {\bibinfo {volume} {49}},\
  \bibinfo {pages} {561} (\bibinfo {year} {2020})}\BibitemShut {NoStop}%
\bibitem [{\citenamefont {Dimitrov}\ \emph {et~al.}(2011)\citenamefont
  {Dimitrov}, \citenamefont {Petrova}, \citenamefont {Kozarova}, \citenamefont
  {Apostolova},\ and\ \citenamefont {Tsvetanov}}]{Dimitrov_SM_2011}%
  \BibitemOpen
  \bibfield  {author} {\bibinfo {author} {\bibfnamefont {I.~V.}\ \bibnamefont
  {Dimitrov}}, \bibinfo {author} {\bibfnamefont {E.~B.}\ \bibnamefont
  {Petrova}}, \bibinfo {author} {\bibfnamefont {R.~G.}\ \bibnamefont
  {Kozarova}}, \bibinfo {author} {\bibfnamefont {M.~D.}\ \bibnamefont
  {Apostolova}}, \ and\ \bibinfo {author} {\bibfnamefont {C.~B.}\ \bibnamefont
  {Tsvetanov}},\ }\href {\doibase 10.1039/C1SM05805C} {\bibfield  {journal}
  {\bibinfo  {journal} {Soft Matter}\ }\textbf {\bibinfo {volume} {7}},\
  \bibinfo {pages} {8002} (\bibinfo {year} {2011})}\BibitemShut {NoStop}%
\bibitem [{\citenamefont {Jour~Putnam}(2006)}]{Putam_Nat_2006}%
  \BibitemOpen
  \bibfield  {author} {\bibinfo {author} {\bibfnamefont {D.}~\bibnamefont
  {Jour~Putnam}},\ }\href {\doibase 10.1038/nmat1645} {\bibfield  {journal}
  {\bibinfo  {journal} {Nature Materials}\ }\textbf {\bibinfo {volume} {5}},\
  \bibinfo {pages} {1476} (\bibinfo {year} {2006})}\BibitemShut {NoStop}%
\bibitem [{\citenamefont {des Rieux}\ \emph {et~al.}(2009)\citenamefont {des
  Rieux}, \citenamefont {Shikanov},\ and\ \citenamefont
  {Shea}}]{Anne_JCR_2009}%
  \BibitemOpen
  \bibfield  {author} {\bibinfo {author} {\bibfnamefont {A.}~\bibnamefont {des
  Rieux}}, \bibinfo {author} {\bibfnamefont {A.}~\bibnamefont {Shikanov}}, \
  and\ \bibinfo {author} {\bibfnamefont {L.~D.}\ \bibnamefont {Shea}},\ }\href
  {\doibase https://doi.org/10.1016/j.jconrel.2009.02.004} {\bibfield
  {journal} {\bibinfo  {journal} {Journal of Controlled Release}\ }\textbf
  {\bibinfo {volume} {136}},\ \bibinfo {pages} {148 } (\bibinfo {year}
  {2009})}\BibitemShut {NoStop}%
\bibitem [{\citenamefont {Haladjova}\ \emph {et~al.}(2012)\citenamefont
  {Haladjova}, \citenamefont {Rangelov}, \citenamefont {Tsvetanov},\ and\
  \citenamefont {Pispas}}]{Haladjova_SM_2012}%
  \BibitemOpen
  \bibfield  {author} {\bibinfo {author} {\bibfnamefont {E.}~\bibnamefont
  {Haladjova}}, \bibinfo {author} {\bibfnamefont {S.}~\bibnamefont {Rangelov}},
  \bibinfo {author} {\bibfnamefont {C.~B.}\ \bibnamefont {Tsvetanov}}, \ and\
  \bibinfo {author} {\bibfnamefont {S.}~\bibnamefont {Pispas}},\ }\href
  {\doibase 10.1039/C2SM07029D} {\bibfield  {journal} {\bibinfo  {journal}
  {Soft Matter}\ }\textbf {\bibinfo {volume} {8}},\ \bibinfo {pages} {2884}
  (\bibinfo {year} {2012})}\BibitemShut {NoStop}%
\bibitem [{\citenamefont {Auyeung}\ \emph {et~al.}(2012)\citenamefont
  {Auyeung}, \citenamefont {Macfarlane}, \citenamefont {Choi}, \citenamefont
  {Cutler},\ and\ \citenamefont {Mirkin}}]{Mirkin_AM_2012}%
  \BibitemOpen
  \bibfield  {author} {\bibinfo {author} {\bibfnamefont {E.}~\bibnamefont
  {Auyeung}}, \bibinfo {author} {\bibfnamefont {R.~J.}\ \bibnamefont
  {Macfarlane}}, \bibinfo {author} {\bibfnamefont {C.~H.~J.}\ \bibnamefont
  {Choi}}, \bibinfo {author} {\bibfnamefont {J.~I.}\ \bibnamefont {Cutler}}, \
  and\ \bibinfo {author} {\bibfnamefont {C.~A.}\ \bibnamefont {Mirkin}},\
  }\href@noop {} {\bibfield  {journal} {\bibinfo  {journal} {Advanced
  Materials}\ }\textbf {\bibinfo {volume} {24}},\ \bibinfo {pages} {5181}
  (\bibinfo {year} {2012})}\BibitemShut {NoStop}%
\bibitem [{\citenamefont {Mirkin}\ \emph {et~al.}(1996)\citenamefont {Mirkin},
  \citenamefont {Letsinger}, \citenamefont {Mucic},\ and\ \citenamefont
  {Storhoff}}]{Mirkin_Nat_1996}%
  \BibitemOpen
  \bibfield  {author} {\bibinfo {author} {\bibfnamefont {C.~A.}\ \bibnamefont
  {Mirkin}}, \bibinfo {author} {\bibfnamefont {R.~L.}\ \bibnamefont
  {Letsinger}}, \bibinfo {author} {\bibfnamefont {R.~C.}\ \bibnamefont
  {Mucic}}, \ and\ \bibinfo {author} {\bibfnamefont {J.~J.}\ \bibnamefont
  {Storhoff}},\ }\href@noop {} {\bibfield  {journal} {\bibinfo  {journal}
  {Nature}\ }\textbf {\bibinfo {volume} {382}},\ \bibinfo {pages} {607}
  (\bibinfo {year} {1996})}\BibitemShut {NoStop}%
\bibitem [{\citenamefont {Kumar}\ \emph {et~al.}(2017)\citenamefont {Kumar},
  \citenamefont {Kumar}, \citenamefont {Giri},\ and\ \citenamefont
  {Nath}}]{Sanjay_EPL_2017}%
  \BibitemOpen
  \bibfield  {author} {\bibinfo {author} {\bibfnamefont {S.}~\bibnamefont
  {Kumar}}, \bibinfo {author} {\bibfnamefont {S.}~\bibnamefont {Kumar}},
  \bibinfo {author} {\bibfnamefont {D.}~\bibnamefont {Giri}}, \ and\ \bibinfo
  {author} {\bibfnamefont {S.}~\bibnamefont {Nath}},\ }\href@noop {} {\bibfield
   {journal} {\bibinfo  {journal} {Europhysics Letters}\ }\textbf {\bibinfo
  {volume} {118}},\ \bibinfo {pages} {28001} (\bibinfo {year}
  {2017})}\BibitemShut {NoStop}%
\bibitem [{\citenamefont {Turner}\ \emph {et~al.}(2002)\citenamefont {Turner},
  \citenamefont {Cabodi},\ and\ \citenamefont {Craighead}}]{Turner_PRL_2002}%
  \BibitemOpen
  \bibfield  {author} {\bibinfo {author} {\bibfnamefont {S.~W.~P.}\
  \bibnamefont {Turner}}, \bibinfo {author} {\bibfnamefont {M.}~\bibnamefont
  {Cabodi}}, \ and\ \bibinfo {author} {\bibfnamefont {H.~G.}\ \bibnamefont
  {Craighead}},\ }\href {\doibase 10.1103/PhysRevLett.88.128103} {\bibfield
  {journal} {\bibinfo  {journal} {Phys. Rev. Lett.}\ }\textbf {\bibinfo
  {volume} {88}},\ \bibinfo {pages} {128103} (\bibinfo {year}
  {2002})}\BibitemShut {NoStop}%
\bibitem [{\citenamefont {Maity}\ \emph {et~al.}(2017)\citenamefont {Maity},
  \citenamefont {Singh},\ and\ \citenamefont {Singh}}]{Maity_EBPJ_2017}%
  \BibitemOpen
  \bibfield  {author} {\bibinfo {author} {\bibfnamefont {A.}~\bibnamefont
  {Maity}}, \bibinfo {author} {\bibfnamefont {A.}~\bibnamefont {Singh}}, \ and\
  \bibinfo {author} {\bibfnamefont {N.}~\bibnamefont {Singh}},\ }\href
  {\doibase 10.1007/s00249-016-1132-3} {\bibfield  {journal} {\bibinfo
  {journal} {European Biophysics Journal}\ }\textbf {\bibinfo {volume} {46}},\
  \bibinfo {pages} {33} (\bibinfo {year} {2017})}\BibitemShut {NoStop}%
\bibitem [{\citenamefont {Akabayov}\ \emph {et~al.}(2013)\citenamefont
  {Akabayov}, \citenamefont {Akabayov}, \citenamefont {Lee}, \citenamefont
  {Wagner},\ and\ \citenamefont {Richardson}}]{Akabayov_NatC_2013}%
  \BibitemOpen
  \bibfield  {author} {\bibinfo {author} {\bibfnamefont {B.}~\bibnamefont
  {Akabayov}}, \bibinfo {author} {\bibfnamefont {S.}~\bibnamefont {Akabayov}},
  \bibinfo {author} {\bibfnamefont {S.}~\bibnamefont {Lee}}, \bibinfo {author}
  {\bibfnamefont {G.}~\bibnamefont {Wagner}}, \ and\ \bibinfo {author}
  {\bibfnamefont {C.}~\bibnamefont {Richardson}},\ }\href@noop {} {\bibfield
  {journal} {\bibinfo  {journal} {Nature Communications}\ }\textbf {\bibinfo
  {volume} {4}} (\bibinfo {year} {2013})}\BibitemShut {NoStop}%
\bibitem [{\citenamefont {Dauxois}\ \emph {et~al.}(1993)\citenamefont
  {Dauxois}, \citenamefont {Peyrard},\ and\ \citenamefont
  {Bishop}}]{Dauxois_PRE_1993}%
  \BibitemOpen
  \bibfield  {author} {\bibinfo {author} {\bibfnamefont {T.}~\bibnamefont
  {Dauxois}}, \bibinfo {author} {\bibfnamefont {M.}~\bibnamefont {Peyrard}}, \
  and\ \bibinfo {author} {\bibfnamefont {A.~R.}\ \bibnamefont {Bishop}},\
  }\href {\doibase 10.1103/PhysRevE.47.R44} {\bibfield  {journal} {\bibinfo
  {journal} {Phys. Rev. E}\ }\textbf {\bibinfo {volume} {47}},\ \bibinfo
  {pages} {R44} (\bibinfo {year} {1993})}\BibitemShut {NoStop}%
\bibitem [{\citenamefont {Frank-Kamenetskii}\ and\ \citenamefont
  {Prakash}(2014)}]{Frank_PLR_2014}%
  \BibitemOpen
  \bibfield  {author} {\bibinfo {author} {\bibfnamefont {M.~D.}\ \bibnamefont
  {Frank-Kamenetskii}}\ and\ \bibinfo {author} {\bibfnamefont {S.}~\bibnamefont
  {Prakash}},\ }\href {\doibase http://dx.doi.org/10.1016/j.plrev.2014.01.005}
  {\bibfield  {journal} {\bibinfo  {journal} {Physics of Life Reviews}\
  }\textbf {\bibinfo {volume} {11}},\ \bibinfo {pages} {153 } (\bibinfo {year}
  {2014})}\BibitemShut {NoStop}%
\bibitem [{\citenamefont {Cocco}\ and\ \citenamefont
  {Monasson}(1999)}]{Cocco_PRL_1999}%
  \BibitemOpen
  \bibfield  {author} {\bibinfo {author} {\bibfnamefont {S.}~\bibnamefont
  {Cocco}}\ and\ \bibinfo {author} {\bibfnamefont {R.}~\bibnamefont
  {Monasson}},\ }\href {\doibase 10.1103/PhysRevLett.83.5178} {\bibfield
  {journal} {\bibinfo  {journal} {Phys. Rev. Lett.}\ }\textbf {\bibinfo
  {volume} {83}},\ \bibinfo {pages} {5178 } (\bibinfo {year}
  {1999})}\BibitemShut {NoStop}%
\bibitem [{\citenamefont {Zoli}(2021)}]{Zoli_JCP_2021}%
  \BibitemOpen
  \bibfield  {author} {\bibinfo {author} {\bibfnamefont {M.}~\bibnamefont
  {Zoli}},\ }\href {\doibase 10.1063/5.0046891} {\bibfield  {journal} {\bibinfo
   {journal} {The Journal of Chemical Physics}\ }\textbf {\bibinfo {volume}
  {154}},\ \bibinfo {pages} {194102} (\bibinfo {year} {2021})}\BibitemShut
  {NoStop}%
\bibitem [{\citenamefont {Zoli}(2019)}]{Zoli_PCCP_2019}%
  \BibitemOpen
  \bibfield  {author} {\bibinfo {author} {\bibfnamefont {M.}~\bibnamefont
  {Zoli}},\ }\href {\doibase 10.1039/C9CP01098J} {\bibfield  {journal}
  {\bibinfo  {journal} {Phys. Chem. Chem. Phys.}\ }\textbf {\bibinfo {volume}
  {21}},\ \bibinfo {pages} {12566} (\bibinfo {year} {2019})}\BibitemShut
  {NoStop}%
\bibitem [{\citenamefont {{Rodrigues Leal}}\ and\ \citenamefont
  {Weber}(2020)}]{Weber_CPL_2020}%
  \BibitemOpen
  \bibfield  {author} {\bibinfo {author} {\bibfnamefont {M.}~\bibnamefont
  {{Rodrigues Leal}}}\ and\ \bibinfo {author} {\bibfnamefont {G.}~\bibnamefont
  {Weber}},\ }\href {\doibase https://doi.org/10.1016/j.cplett.2020.137781}
  {\bibfield  {journal} {\bibinfo  {journal} {Chemical Physics Letters}\
  }\textbf {\bibinfo {volume} {755}},\ \bibinfo {pages} {137781} (\bibinfo
  {year} {2020})}\BibitemShut {NoStop}%
\bibitem [{\citenamefont {Singh}\ and\ \citenamefont
  {Singh}(2005)}]{Navin_epje_2005}%
  \BibitemOpen
  \bibfield  {author} {\bibinfo {author} {\bibfnamefont {N.}~\bibnamefont
  {Singh}}\ and\ \bibinfo {author} {\bibfnamefont {Y.}~\bibnamefont {Singh}},\
  }\href {\doibase 10.1140/epje/i2004-10100-7} {\bibfield  {journal} {\bibinfo
  {journal} {The European Physical Journal E}\ }\textbf {\bibinfo {volume}
  {17}},\ \bibinfo {pages} {7} (\bibinfo {year} {2005})}\BibitemShut {NoStop}%
\bibitem [{\citenamefont {Peyrard}\ and\ \citenamefont
  {Bishop}(1989)}]{Peyrard_PRL_1989}%
  \BibitemOpen
  \bibfield  {author} {\bibinfo {author} {\bibfnamefont {M.}~\bibnamefont
  {Peyrard}}\ and\ \bibinfo {author} {\bibfnamefont {A.~R.}\ \bibnamefont
  {Bishop}},\ }\href {\doibase 10.1103/PhysRevLett.62.2755} {\bibfield
  {journal} {\bibinfo  {journal} {Phys. Rev. Lett.}\ }\textbf {\bibinfo
  {volume} {62}},\ \bibinfo {pages} {2755} (\bibinfo {year}
  {1989})}\BibitemShut {NoStop}%
\bibitem [{\citenamefont {Cule}\ and\ \citenamefont
  {Hwa}(1997)}]{Cule_PRL_1997}%
  \BibitemOpen
  \bibfield  {author} {\bibinfo {author} {\bibfnamefont {D.}~\bibnamefont
  {Cule}}\ and\ \bibinfo {author} {\bibfnamefont {T.}~\bibnamefont {Hwa}},\
  }\href {\doibase 10.1103/PhysRevLett.79.2375} {\bibfield  {journal} {\bibinfo
   {journal} {Phys. Rev. Lett.}\ }\textbf {\bibinfo {volume} {79}},\ \bibinfo
  {pages} {2375} (\bibinfo {year} {1997})}\BibitemShut {NoStop}%
\bibitem [{\citenamefont {Zhang}\ \emph {et~al.}(1997)\citenamefont {Zhang},
  \citenamefont {Zheng}, \citenamefont {Liu},\ and\ \citenamefont
  {Chen}}]{Zhang_PRE_1997}%
  \BibitemOpen
  \bibfield  {author} {\bibinfo {author} {\bibfnamefont {Y.-l.}\ \bibnamefont
  {Zhang}}, \bibinfo {author} {\bibfnamefont {W.-M.}\ \bibnamefont {Zheng}},
  \bibinfo {author} {\bibfnamefont {J.-X.}\ \bibnamefont {Liu}}, \ and\
  \bibinfo {author} {\bibfnamefont {Y.~Z.}\ \bibnamefont {Chen}},\ }\href
  {\doibase 10.1103/PhysRevE.56.7100} {\bibfield  {journal} {\bibinfo
  {journal} {Phys. Rev. E}\ }\textbf {\bibinfo {volume} {56}},\ \bibinfo
  {pages} {7100} (\bibinfo {year} {1997})}\BibitemShut {NoStop}%
\bibitem [{\citenamefont {Campa}\ and\ \citenamefont
  {Giansanti}(1998)}]{Campa_PRE_1998}%
  \BibitemOpen
  \bibfield  {author} {\bibinfo {author} {\bibfnamefont {A.}~\bibnamefont
  {Campa}}\ and\ \bibinfo {author} {\bibfnamefont {A.}~\bibnamefont
  {Giansanti}},\ }\href {\doibase 10.1103/PhysRevE.58.3585} {\bibfield
  {journal} {\bibinfo  {journal} {Phys. Rev. E}\ }\textbf {\bibinfo {volume}
  {58}},\ \bibinfo {pages} {3585} (\bibinfo {year} {1998})}\BibitemShut
  {NoStop}%
\bibitem [{\citenamefont {van Erp}\ \emph {et~al.}(2006)\citenamefont {van
  Erp}, \citenamefont {Cuesta-Lopez},\ and\ \citenamefont
  {Peyrard}}]{Erp_epje_2006}%
  \BibitemOpen
  \bibfield  {author} {\bibinfo {author} {\bibfnamefont {T.~S.}\ \bibnamefont
  {van Erp}}, \bibinfo {author} {\bibfnamefont {S.}~\bibnamefont
  {Cuesta-Lopez}}, \ and\ \bibinfo {author} {\bibfnamefont {M.}~\bibnamefont
  {Peyrard}},\ }\href {\doibase 10.1140/epje/i2006-10032-2} {\bibfield
  {journal} {\bibinfo  {journal} {The European Physical Journal E}\ }\textbf
  {\bibinfo {volume} {20}},\ \bibinfo {pages} {421} (\bibinfo {year}
  {2006})}\BibitemShut {NoStop}%
\bibitem [{\citenamefont {Dauxois}\ and\ \citenamefont
  {Peyrard}(1995)}]{Dauxois_PRE_1995}%
  \BibitemOpen
  \bibfield  {author} {\bibinfo {author} {\bibfnamefont {T.}~\bibnamefont
  {Dauxois}}\ and\ \bibinfo {author} {\bibfnamefont {M.}~\bibnamefont
  {Peyrard}},\ }\href {\doibase 10.1103/PhysRevE.51.4027} {\bibfield  {journal}
  {\bibinfo  {journal} {Phys. Rev. E}\ }\textbf {\bibinfo {volume} {51}},\
  \bibinfo {pages} {4027} (\bibinfo {year} {1995})}\BibitemShut {NoStop}%
\bibitem [{\citenamefont {Singh}\ and\ \citenamefont
  {Singh}(2015)}]{Amar_PRE_2015}%
  \BibitemOpen
  \bibfield  {author} {\bibinfo {author} {\bibfnamefont {A.}~\bibnamefont
  {Singh}}\ and\ \bibinfo {author} {\bibfnamefont {N.}~\bibnamefont {Singh}},\
  }\href {\doibase 10.1103/PhysRevE.92.032703} {\bibfield  {journal} {\bibinfo
  {journal} {Phys. Rev. E}\ }\textbf {\bibinfo {volume} {92}},\ \bibinfo
  {pages} {032703} (\bibinfo {year} {2015})}\BibitemShut {NoStop}%
\bibitem [{\citenamefont {Lavery}\ \emph {et~al.}(2009)\citenamefont {Lavery},
  \citenamefont {Moakher}, \citenamefont {Maddocks}, \citenamefont
  {Petkeviciute},\ and\ \citenamefont {Zakrzewska}}]{Lavery_NAR_2009}%
  \BibitemOpen
  \bibfield  {author} {\bibinfo {author} {\bibfnamefont {R.}~\bibnamefont
  {Lavery}}, \bibinfo {author} {\bibfnamefont {M.}~\bibnamefont {Moakher}},
  \bibinfo {author} {\bibfnamefont {J.~H.}\ \bibnamefont {Maddocks}}, \bibinfo
  {author} {\bibfnamefont {D.}~\bibnamefont {Petkeviciute}}, \ and\ \bibinfo
  {author} {\bibfnamefont {K.}~\bibnamefont {Zakrzewska}},\ }\href {\doibase
  10.1093/nar/gkp608} {\bibfield  {journal} {\bibinfo  {journal} {Nucleic Acids
  Research}\ }\textbf {\bibinfo {volume} {37}},\ \bibinfo {pages} {5917}
  (\bibinfo {year} {2009})}\BibitemShut {NoStop}%
\bibitem [{\citenamefont {Olson}\ \emph {et~al.}(2001)\citenamefont {Olson},
  \citenamefont {Bansal}, \citenamefont {Burley}, \citenamefont {Dickerson},
  \citenamefont {Gerstein}, \citenamefont {Harvey}, \citenamefont {Heinemann},
  \citenamefont {Lu}, \citenamefont {Neidle}, \citenamefont {Shakked},
  \citenamefont {Sklenar}, \citenamefont {Suzuki}, \citenamefont {Tung},
  \citenamefont {Westhof}, \citenamefont {Wolberger},\ and\ \citenamefont
  {Berman}}]{Wilma_JMB_2001}%
  \BibitemOpen
  \bibfield  {author} {\bibinfo {author} {\bibfnamefont {W.~K.}\ \bibnamefont
  {Olson}}, \bibinfo {author} {\bibfnamefont {M.}~\bibnamefont {Bansal}},
  \bibinfo {author} {\bibfnamefont {S.~K.}\ \bibnamefont {Burley}}, \bibinfo
  {author} {\bibfnamefont {R.~E.}\ \bibnamefont {Dickerson}}, \bibinfo {author}
  {\bibfnamefont {M.}~\bibnamefont {Gerstein}}, \bibinfo {author}
  {\bibfnamefont {S.~C.}\ \bibnamefont {Harvey}}, \bibinfo {author}
  {\bibfnamefont {U.}~\bibnamefont {Heinemann}}, \bibinfo {author}
  {\bibfnamefont {X.-J.}\ \bibnamefont {Lu}}, \bibinfo {author} {\bibfnamefont
  {S.}~\bibnamefont {Neidle}}, \bibinfo {author} {\bibfnamefont
  {Z.}~\bibnamefont {Shakked}}, \bibinfo {author} {\bibfnamefont
  {H.}~\bibnamefont {Sklenar}}, \bibinfo {author} {\bibfnamefont
  {M.}~\bibnamefont {Suzuki}}, \bibinfo {author} {\bibfnamefont {C.-S.}\
  \bibnamefont {Tung}}, \bibinfo {author} {\bibfnamefont {E.}~\bibnamefont
  {Westhof}}, \bibinfo {author} {\bibfnamefont {C.}~\bibnamefont {Wolberger}},
  \ and\ \bibinfo {author} {\bibfnamefont {H.~M.}\ \bibnamefont {Berman}},\
  }\href {\doibase https://doi.org/10.1006/jmbi.2001.4987} {\bibfield
  {journal} {\bibinfo  {journal} {Journal of Molecular Biology}\ }\textbf
  {\bibinfo {volume} {313}},\ \bibinfo {pages} {229 } (\bibinfo {year}
  {2001})}\BibitemShut {NoStop}%
\bibitem [{\citenamefont {Bansal}\ and\ \citenamefont
  {Ravi}(1995)}]{Bansal_CAB_1995}%
  \BibitemOpen
  \bibfield  {author} {\bibinfo {author} {\bibfnamefont {D.~B.}\ \bibnamefont
  {Bansal}, \bibfnamefont {M.}}\ and\ \bibinfo {author} {\bibfnamefont
  {B.}~\bibnamefont {Ravi}},\ }\href@noop {} {\bibfield  {journal} {\bibinfo
  {journal} {Comput Appl Biosci}\ }\textbf {\bibinfo {volume} {11}},\ \bibinfo
  {pages} {281} (\bibinfo {year} {1995})}\BibitemShut {NoStop}%
\bibitem [{\citenamefont {Bevan}\ \emph {et~al.}(2000)\citenamefont {Bevan},
  \citenamefont {Li}, \citenamefont {Pedersen},\ and\ \citenamefont
  {Darden}}]{Bevan_BJ_2000}%
  \BibitemOpen
  \bibfield  {author} {\bibinfo {author} {\bibfnamefont {D.~R.}\ \bibnamefont
  {Bevan}}, \bibinfo {author} {\bibfnamefont {L.}~\bibnamefont {Li}}, \bibinfo
  {author} {\bibfnamefont {L.~G.}\ \bibnamefont {Pedersen}}, \ and\ \bibinfo
  {author} {\bibfnamefont {T.~A.}\ \bibnamefont {Darden}},\ }\href {\doibase
  https://doi.org/10.1016/S0006-3495(00)76625-2} {\bibfield  {journal}
  {\bibinfo  {journal} {Biophysical Journal}\ }\textbf {\bibinfo {volume}
  {78}},\ \bibinfo {pages} {668 } (\bibinfo {year} {2000})}\BibitemShut
  {NoStop}%
\bibitem [{\citenamefont {Samanta}\ \emph {et~al.}(2009)\citenamefont
  {Samanta}, \citenamefont {Mukherjee}, \citenamefont {Chakrabarti},\ and\
  \citenamefont {Bhattacharyya}}]{Dhananjay_JCP_2009}%
  \BibitemOpen
  \bibfield  {author} {\bibinfo {author} {\bibfnamefont {S.}~\bibnamefont
  {Samanta}}, \bibinfo {author} {\bibfnamefont {S.}~\bibnamefont {Mukherjee}},
  \bibinfo {author} {\bibfnamefont {J.}~\bibnamefont {Chakrabarti}}, \ and\
  \bibinfo {author} {\bibfnamefont {D.}~\bibnamefont {Bhattacharyya}},\ }\href
  {\doibase 10.1063/1.3078797} {\bibfield  {journal} {\bibinfo  {journal} {The
  Journal of Chemical Physics}\ }\textbf {\bibinfo {volume} {130}},\ \bibinfo
  {pages} {115103} (\bibinfo {year} {2009})}\BibitemShut {NoStop}%
\bibitem [{\citenamefont {Cheng}\ \emph {et~al.}(2006)\citenamefont {Cheng},
  \citenamefont {Korolev},\ and\ \citenamefont
  {Nordenskiöld}}]{Cheng_NAR_2006}%
  \BibitemOpen
  \bibfield  {author} {\bibinfo {author} {\bibfnamefont {Y.}~\bibnamefont
  {Cheng}}, \bibinfo {author} {\bibfnamefont {N.}~\bibnamefont {Korolev}}, \
  and\ \bibinfo {author} {\bibfnamefont {L.}~\bibnamefont {Nordenskiöld}},\
  }\href {\doibase 10.1093/nar/gkj434} {\bibfield  {journal} {\bibinfo
  {journal} {Nucleic Acids Research}\ }\textbf {\bibinfo {volume} {34}},\
  \bibinfo {pages} {686} (\bibinfo {year} {2006})}\BibitemShut {NoStop}%
\bibitem [{\citenamefont {Dai}\ \emph {et~al.}(2008)\citenamefont {Dai},
  \citenamefont {Mu}, \citenamefont {Nordenski\"old},\ and\ \citenamefont
  {van~der Maarel}}]{Dai_PRL_2008}%
  \BibitemOpen
  \bibfield  {author} {\bibinfo {author} {\bibfnamefont {L.}~\bibnamefont
  {Dai}}, \bibinfo {author} {\bibfnamefont {Y.}~\bibnamefont {Mu}}, \bibinfo
  {author} {\bibfnamefont {L.}~\bibnamefont {Nordenski\"old}}, \ and\ \bibinfo
  {author} {\bibfnamefont {J.~R.~C.}\ \bibnamefont {van~der Maarel}},\ }\href
  {\doibase 10.1103/PhysRevLett.100.118301} {\bibfield  {journal} {\bibinfo
  {journal} {Phys. Rev. Lett.}\ }\textbf {\bibinfo {volume} {100}},\ \bibinfo
  {pages} {118301} (\bibinfo {year} {2008})}\BibitemShut {NoStop}%
\bibitem [{\citenamefont {Long}\ \emph {et~al.}(2006)\citenamefont {Long},
  \citenamefont {Kudlay},\ and\ \citenamefont {Schatz}}]{Long_JPC_2006}%
  \BibitemOpen
  \bibfield  {author} {\bibinfo {author} {\bibfnamefont {H.}~\bibnamefont
  {Long}}, \bibinfo {author} {\bibfnamefont {A.}~\bibnamefont {Kudlay}}, \ and\
  \bibinfo {author} {\bibfnamefont {G.~C.}\ \bibnamefont {Schatz}},\ }\href
  {\doibase 10.1021/jp0556815} {\bibfield  {journal} {\bibinfo  {journal} {The
  Journal of Physical Chemistry B}\ }\textbf {\bibinfo {volume} {110}},\
  \bibinfo {pages} {2918} (\bibinfo {year} {2006})}\BibitemShut {NoStop}%
\bibitem [{\citenamefont {Olson}\ \emph {et~al.}(1998)\citenamefont {Olson},
  \citenamefont {Gorin}, \citenamefont {Lu}, \citenamefont {Hock},\ and\
  \citenamefont {Zhurkin}}]{Olson_PNAS_1998}%
  \BibitemOpen
  \bibfield  {author} {\bibinfo {author} {\bibfnamefont {W.~K.}\ \bibnamefont
  {Olson}}, \bibinfo {author} {\bibfnamefont {A.~A.}\ \bibnamefont {Gorin}},
  \bibinfo {author} {\bibfnamefont {X.-J.}\ \bibnamefont {Lu}}, \bibinfo
  {author} {\bibfnamefont {L.~M.}\ \bibnamefont {Hock}}, \ and\ \bibinfo
  {author} {\bibfnamefont {V.~B.}\ \bibnamefont {Zhurkin}},\ }\href {\doibase
  10.1073/pnas.95.19.11163} {\bibfield  {journal} {\bibinfo  {journal}
  {Proceedings of the National Academy of Sciences}\ }\textbf {\bibinfo
  {volume} {95}},\ \bibinfo {pages} {11163} (\bibinfo {year}
  {1998})}\BibitemShut {NoStop}%
\end{thebibliography}%

\end{document}